\documentclass[aps,pra,twocolumn,superscriptaddress]{revtex4-1}

\usepackage{graphicx}
\usepackage{amssymb}
\usepackage{amsmath}
\usepackage{color}
\usepackage[dvipdfm]{hyperref}

\begin{document}

\title{Emergent Coulombic criticality and Kibble-Zurek scaling in 
a topological magnet} 

\author{James Hamp}
\email[]{joh28@cam.ac.uk}
\affiliation{TCM Group, Cavendish Laboratory, University of Cambridge, J.~J.~Thomson Avenue, Cambridge CB3 0HE, United Kingdom}

\author{Anushya Chandran}
\affiliation{Perimeter Institute for Theoretical Physics, 31 Caroline Street North, Waterloo, Ontario N2L 2Y5, Canada}

\author{Roderich Moessner}
\affiliation{Max-Planck-Institut f\"ur Physik komplexer Systeme, N\"othnitzer Stra{\ss}e 38, Dresden 01187, Germany}

\author{Claudio Castelnovo}
\affiliation{TCM Group, Cavendish Laboratory, University of Cambridge, J.~J.~Thomson Avenue, Cambridge CB3 0HE, United Kingdom}

\date{\today}

\begin{abstract}
When a classical system is driven through a continuous phase transition, its nonequilibrium response is universal and exhibits Kibble-Zurek scaling.
We explore this dynamical scaling in the novel context of a three-dimensional topological magnet with fractionalized excitations, namely the liquid-gas transition of the emergent mobile magnetic monopoles
in dipolar spin ice.
Using field-mixing and finite-size scaling techniques, we place the critical point of the liquid-gas line in the three-dimensional Ising universality class.  
We then demonstrate Kibble-Zurek scaling for sweeps of the magnetic field through the critical point. 
Unusually slow microscopic time scales in spin ice offer a unique opportunity to detect this universal nonequilibrium physics in current experimental setups. 
\end{abstract}


\maketitle

\section{Introduction}

The theory of equilibrium phase transitions is one of the major achievements of 20th-century physics~\cite{goldenfeld_lectures_1992}. 
Effective theories based on local order parameters describe the universal aspects of such transitions. 
Our understanding of nonequilibrium physics in their vicinity has also progressed on many fronts, from scaling properties of dynamical correlation functions~\cite{Hohenberg:1977aa,goldenfeld_lectures_1992}, to Kibble-Zurek (KZ) behavior~\cite{Dziarmaga:2010qf, polkovnikov_colloquium:_2011, del_campo_universality_2014}. 

Topological phases do not fit into the conventional framework as they are not characterized by local order parameters~\cite{Wen_book_2004}. 
Indeed, one of the principal attractions of systems with emergent gauge fields is that standard Landau-Ginzburg-Wilson arguments can fail. 
The phenomenon of deconfined quantum criticality~\cite{senthil_deconfined_2004} and analogous phenomena noted earlier in quantum dimer models~\cite{moessner_sondhi_fradkin_2001} are examples of this.
On general grounds this suggests that critical points can exhibit new and unexpected universality classes, and it is not clear in which settings this happens. 

This work combines the study of a critical point in a topological system with an investigation of topological phases out of equilibrium. This is a young, multi-faceted and rapidly developing subject~\footnote{For a collection of current topics, see e.g., \href{http://www.pks.mpg.de/\textasciitilde tomaeq14/}{http://www.pks.mpg.de/\textasciitilde tomaeq14/}}. 
In particular, the interplay of nonequilibrium physics, long-range interactions and classical topological order near a critical point is largely unchartered territory, which this work explores. 

We consider three-dimensional dipolar spin ice, realized in rare-earth pyrochlore oxides such as $\{$Dy,Ho$\}$$_2$Ti$_2$O$_7$. 
At low temperatures, frustration prevents the system from ordering~\cite{bramwell_spin_2001} and it enters a highly degenerate topological regime described by an emergent gauge field  with magnetic Coulomb-interacting monopoles as fractional excitations~\cite{castelnovo_spin_2012}. 
In many---primarily thermodynamic---respects, spin ice is well modeled as a magnetic version of an electrolyte~\cite{castelnovo_DH_2011}, whereas---particularly out of equilibrium---fundamental deviations on account of the Dirac strings connecting the monopoles have been observed~\cite{castelnovo_thquench_2010,kaiser_ac_2014}.

In an appropriately oriented magnetic field, spin ice exhibits a liquid-gas phase diagram in the temperature-field plane, where a first-order transition line terminates at a critical point~\cite{sakakibara_observation_2003}. 
This behavior is extremely unusual for a system of localized spins. 
However, its origin is naturally understood in the monopole picture: the field acts as a tunable chemical potential for the monopoles, which form a type of Coulomb liquid~\cite{castelnovo_magnetic_2008}, well known to exhibit liquid-gas phase diagrams~\cite{fisher_story_1994}.

Our study combines the above questions on critical and nonequilibrium properties of topological systems by, 
firstly, characterizing this critical point and, secondly, studying the universal dynamics in its vicinity.

We characterize equilibrium properties via field mixing analysis~\cite{wilding_simulation_1997} and finite-size scaling. 
This analysis strongly suggests that the critical point is in the three-dimensional (3D) Ising universality class, as is believed to be the case for conventional Coulomb liquids~\cite{fisher_story_1994, luijten_universality_2002}. 

We then analyze local (single spin-flip) dynamics  in the vicinity of this transition. 
We study the magnetization response on linearly sweeping the system across the critical point. 
The response is hysteretic and, for slow enough sweeps, universal \`{a} la Kibble-Zurek~\cite{kibble_topology_1976, zurek_cosmological_1985,Dziarmaga:2010qf}. 
We establish the presence of KZ scaling~\cite{Polkovnikov2005, Deng-S.:2008aa, biroli_kibble-zurek_2010, chandran_kibble-zurek_2012}, confirming Ising universality and allowing us to obtain the dynamical scaling exponent. 

Although there is some experimental evidence for KZ scaling of defect density~\cite{zurek_cosmological_1985,del_campo_universality_2014}, a decisive test of the scaling of dynamical response functions in this context is still lacking. 
Our simulations suggest that KZ scaling in spin ice is realistically accessible in field sweep experiments~\cite{slobinsky_unconventional_2010}, owing to unusually slow microscopic time scales of the large rare-earth moments~\cite{snyder_low-temperature_2004,jaubert_signature_2009}. 
In addition,  the (uniform) magnetization turns out to be a direct measure of the monopole density near the critical point---an informative quantity that is otherwise difficult to access experimentally. 
As such, the liquid-gas transition is a classical instance of destabilizing the emergent magnetic vacuum via the ``Schwinger mechanism'' of monopole-antimonopole pair creation~\cite{Schwinger1951}.

Spin ice may thus be a unique experimental system, not only in which to observe this KZ scaling, but also as an instance of a 3D topological material out of equilibrium. 

\section{Field mixing at the critical point} 

We consider dipolar spin ice in a magnetic field of strength $H$ along the [111] direction. 
We use the Hamiltonian 
\begin{align}
	\label{eq:hamiltonian}
\mathcal{H} 
&= 
- \mu H \sum_i \left( \hat{\mathbf{H}} \cdot \hat{\mathbf{e}}_i \right) \sigma_i \;+\; J \sum_{\langle i j \rangle} \sigma_i \sigma_j 
\nonumber \\ 
&+ D r_\mathrm{nn}^3 \sum_{j>i}\left( \frac{\hat{\mathbf{e}}_i \cdot \hat{\mathbf{e}}_j}{|\mathbf{r}_{ij}|^3} - \frac{3(\hat{\mathbf{e}}_i \cdot \mathbf{r}_{ij})(\hat{\mathbf{e}}_j \cdot \mathbf{r}_{ij})}{|\mathbf{r}_{ij}|^5}\right) \sigma_i \sigma_j 
\, , 
\end{align}
where $\{ \sigma_i = \pm 1\}$ are pseudospins of local easy axis $\hat{\mathbf{e}}_i$ and magnetic moment $\mu=10 \, \mu_\mathrm{B}$; $J=-1.24$~K and $D=1.41$~K are respectively the nearest-neighbour exchange and dipolar coupling constants relevant for Dy$_2$Ti$_2$O$_7$ \cite{den_hertog_dipolar_2000}; $r_\mathrm{nn}$ is the nearest-neighbour spacing of the pyrochlore lattice;  $\mathbf{r}_{ij}$ is the separation vector between sites $i$ and $j$, and $\mathbf{H} \equiv H \hat{\mathbf{H}}$ is the external magnetic field which we measure in Tesla (T). 
We set $k_\mathrm{B}=1$ and measure all energies in Kelvin (K).
\begin{figure}[t!]
\centering
\includegraphics[width=1.0\linewidth]{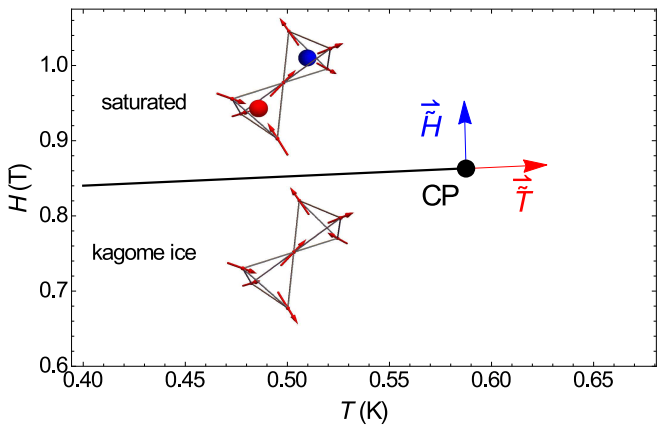}
\caption{\label{fig:phase_diagram}(Color online) Liquid-gas phase diagram of spin ice in the $(T,H)$ plane. 
The insets show representative spin configurations of two tetrahedra in each of the phases. 
The field- and temperaturelike directions at the critical point, $\vec{\tilde{H}}$ and $\vec{\tilde{T}}$, 
are shown. 
Note that the crossover from kagome ice to spin ice occurs outside the range of parameters shown.
}
\end{figure}

The schematic phase diagram of spin ice in the $(T,H)$ plane is shown in Fig.~\ref{fig:phase_diagram}. 
At low $T$, upon increasing the field strength $H$, spin ice crosses over to a partially polarized phase known as kagome ice, and then undergoes a first-order transition to a saturated phase. 
In kagome ice, monopole excitations are activated and exponentially sparse at low $T$, whereas in the saturated phase they are dense as they correspond to the lowest-energy spin configuration. 
$H$ acts as a chemical potential and the transition is characterized by a finite jump in monopole density, from a monopole gas to a monopole liquid phase~\cite{sakakibara_observation_2003,castelnovo_magnetic_2008}. 
Representative spin configurations of two tetrahedra in each of the phases can be seen in the insets of Fig.~\ref{fig:phase_diagram}. 

The first-order line terminates at a critical point, typical of liquid-gas phase diagrams. 
To determine its location, it is convenient to introduce the reduced Hamiltonian 
\begin{equation}
\frac{\mathcal{H}}{T} 
\equiv 
- \beta \varepsilon  
- h m, 
\label{eq: reduced hamiltonian}
\end{equation}
where $\beta$ and $h$ are dimensionless couplings; $m$ is the dimensionless [111] component of the magnetization; and $\varepsilon$ is the dimensionless total spin interaction energy.
The location of the critical point in the $(T,H)$ plane is determined from the crossing of the Binder cumulant of the magnetization $m$ for different system sizes $L$ (see Appendix \ref{app:binder} for details). 
Our best estimate for the critical point thus obtained is $(T_c,H_c) =$ \mbox{$(0.5875 \pm 0.0005 \, \mathrm{K}, 0.86295 \pm 0.00005 \, \mathrm{T})$}.

In liquid-gas phase diagrams, the fieldlike and temperaturelike directions at the critical point typically do not coincide with the original parameters in the system ($H$ and $T$ here). 
The field mixing formalism~\cite{rehr_revised_1973, wilding_simulation_1997}, 
developed in the context of classical fluids, allows the universal properties of such critical points to be extracted. 
According to the revised scaling hypothesis, the correct scaling operators $\tilde{m}$ and $\tilde{\varepsilon}$ are linear superpositions of the quantities $m$ and $\varepsilon$ that appear in the Hamiltonian, 
\begin{equation}
\begin{pmatrix} \tilde{m} \\ \tilde{\varepsilon} \end{pmatrix}  
= 
\begin{pmatrix} 1 & s \\ r & 1 \end{pmatrix} \begin{pmatrix} m \\ \varepsilon \end{pmatrix},
\end{equation}
where $r$ and $s$ are appropriate mixing parameters~\cite{wilding_simulation_1997}. 
The scaling fields conjugate to $\tilde{m}$ and $\tilde{\varepsilon}$---$\tilde{H}$ and $\tilde{T}$ respectively---define new fieldlike (symmetry-breaking) and temperaturelike (non-symmetry-breaking) directions in parameter space. 
The symmetry of the critical point and the universal scaling content of the transition is manifest in the mixed operators and fields.

From the locations of the Binder cumulant minima (alternatively, the maxima of the susceptibilites), the slope of the first-order line in the vicinity of the critical point is determined. 
This slope in the $(\beta,h)$ plane, $\left.\left( \mathrm{d} \beta_c/ \mathrm{d} h \right)\right\vert_{h=h_c}$, is directly related to the mixing parameter $r = ( \mathrm{d} \beta_c / \mathrm{d} h)^{-1} = 4.283 \pm 0.005$.
It is interesting to note that this value is in reasonable agreement with the estimate obtained from the Clausius-Clapeyron relation, $\mathrm{d} H_c / \mathrm{d} T = - \Delta S / \Delta M \simeq 4.45$~\cite{hiroi_specific_2003}, where $\Delta M$ and $\Delta S$ are, respectively, the differences in magnetization and entropy between the kagome ice and saturated phases.
The relation is expected to hold at low $T$ where the transition is strongly first order and has indeed already been shown to extrapolate up to the critical point in \onlinecite{castelnovo_magnetic_2008}.

Obtaining the mixing parameter $s$ is not as straightforward~\cite{rummukainen_universality_1998}. 
Here we do so by requiring the statistical independence of fluctuations in $\tilde{m}$ and $\tilde{\varepsilon}$ at the critical point, and find $s = 0.0(3) \pm 0.0(7)$. 
Other methods give consistent results (not shown). 

A substantial admixture of $\tilde{m}$ in $\varepsilon$ is expected from the monopole picture. 
Indeed, the spin interaction energy in the Hamiltonian encompasses both a Coulombic term as well as a chemical potential for the monopoles, and they are of comparable strength. 
While it is possible to separate (approximately) the two contributions in $\varepsilon$, we find that it does not lead to an appreciable improvement in the field-mixing analysis. 

\begin{figure}[t!]
\centering
\includegraphics[width=1.0\linewidth]{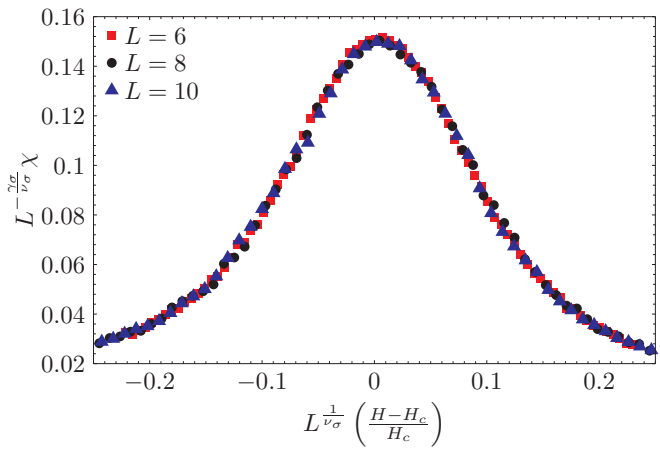}
\caption{\label{fig:collapse_suscep}(Color online) Finite-size scaling collapse of the equilibrium susceptibility $\chi$ in the vicinity of the critical point, using the critical exponents $\gamma_\sigma$ and $\nu_\sigma$ as fitting parameters. The collapse gives \mbox{$\gamma_\sigma = 0.76 \pm 0.02$} and \mbox{$\nu_\sigma = 0.41 \pm 0.01$}, in agreement with the Ising values \mbox{$\gamma_\sigma \simeq 0.79$} and \mbox{$\nu_\sigma \simeq 0.40$}. 
}
\end{figure}

Performing a finite-size scaling analysis in the vicinity of the critical point we obtain critical exponents that are consistent with 3D Ising universality: $\gamma_\sigma = 0.76 \pm 0.02$ and $\nu_\sigma = 0.41 \pm 0.01$, compared with the Ising values $\gamma_\sigma \simeq 0.79$ and $\nu_\sigma \simeq 0.40$ in nonzero field \cite{pelissetto_critical_2002}. 
The finite-size scaling collapse for the magnetic susceptibility $\chi$ is illustrated in Fig.~\ref{fig:collapse_suscep}. 
In the 3D Ising class, fluctuations in $\tilde{m}$ are dominant over those in $\tilde{\varepsilon}$ and we observe the same magnetic susceptibility exponent $\gamma_\sigma$ whether we consider fluctuations in $\tilde{m}$, $m$, or even $\varepsilon$. 
In order to observe the heat capacity exponent $\alpha_{(\sigma)}$, substantial statistical accuracy would be required to ensure that $\tilde{\varepsilon}$ does not contain any contribution from $\tilde{m}$. 

\begin{figure}[t!]
\includegraphics[width=1.0\linewidth]{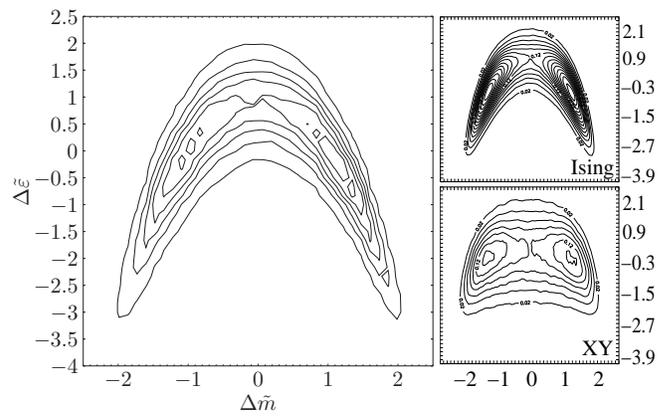}
\caption{\label{fig:banana}(Color online) Histogram of mixed operator fluctuations $\Delta \tilde{m}$, $\Delta \tilde{\varepsilon}$, 
where $\Delta X \equiv X- \langle X \rangle$, normalized by their respective standard deviations, $\sqrt{\langle (\Delta X )^2 \rangle}$. 
The characteristic shape is indicative of emergent $\mathbb{Z}_2$ symmetry and Ising universality. 
Inset: Corresponding data for the 3D Ising and 3D XY models at criticality (adapted from Ref.~\cite{rummukainen_universality_1998}). 
}
\end{figure}

The joint probability distribution of the fluctuations in $\tilde{m}$ and $\tilde{\varepsilon}$ at the critical point, the form of which is known to be universal~\cite{bruce_probability_1981, plascak_probability_2013}, is another powerful tool to identify the universality class of a system~\cite{wilding_liquidvapor_1995,wilding_simulation_1997, rummukainen_universality_1998, karsch_three-dimensional_2000}. 
The distribution can be seen in Fig.~\ref{fig:banana}. It displays a characteristic shape which is the hallmark of an emergent $\mathbb{Z}_2$ symmetry and thus Ising universality. 
The data can be compared with the 3D Ising model ($\mathbb{Z}_2$ symmetry) and contrasted with the 3D XY model [$U(1)$ symmetry] in the insets of Fig.~\ref{fig:banana}. 
Direct comparison between Ising and XY criticality in three dimensions is important due to the very similar scaling exponents, which are difficult to differentiate between within our numerical accuracy. 

Recently, the long-standing question~\cite{fisher_story_1994} of the universality class of critical classical Coulomb liquids has been resolved in numerical simulations~\cite{luijten_universality_2002} in favour of Ising (and not mean-field) behavior. 
Our results are consistent with this, in accordance with the picture of a liquid-gas transition of the emergent monopoles.
It is important to note that the Ising universality does not arise trivially from the Ising nature of the original spins. 
Indeed, their $\mathbb{Z}_2$ symmetry is explicitly broken by the applied field.
The spins constitute a vacuum for the magnetic monopole excitations. It is then this emergent Coulomb liquid, or magnetolyte, which undergoes a liquid-gas transition as would a classical electrolyte, with associated \emph{emergent} Coulombic criticality. 
This is highly nontrivial: emergent monopoles are connected by a network of Dirac strings. These strings are however statistically and energetically immaterial, thus allowing the thermodynamic properties of the system to be the same as for a gas of real pointlike charges. 

\section{Kibble-Zurek scaling of hysteresis}

The standard KZ choice is to vary the temperaturelike parameter 
identified above (see e.g., the recent study \onlinecite{liu_dynamic_2014}). 
However, in the context of a liquid-gas transition, fine tuning would be necessary to identify the precise $\tilde{T}(T,H)$ trajectory, which is in general rather difficult away from the limit $\vert H-H_c \vert ,\,\vert T-T_c \vert \rightarrow 0$. 
As the simplest experimental prototcol, we therefore propose to vary the applied field $H(t)$, which is then a combination of $\tilde{H}$ and $\tilde{T}$, and measure the magnetization $m(t)$. 
The resulting behavior is controlled by the most relevant correlation length in the vicinity of the critical point (the magnetic correlation length in the case of 3D Ising universality, as discussed below). 

\begin{figure}[t!]
\includegraphics[width=1.0\linewidth]{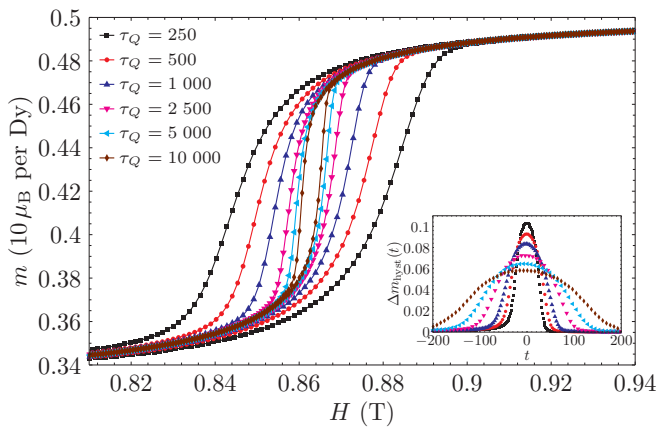}
\caption{\label{fig:bubble_field}(Color online) Hysteresis in the magnetization $m$ during a forwards-backwards field cycle. 
The field $H$ is swept at constant temperature from $0.75$~T, through the critical point, to $0.95$~T in time $\tau_Q$ (measured in MC steps per spin), returning similarly. 
The system size is $L=10$.
Inset: 
Hysteresis loop size $\Delta m_\mathrm{hyst}(t)$ as a function of time to the critical point $t$. 
}
\end{figure}

Let $H(t)$ be varied linearly from $H_i$ to $H_f$ in a time $\tau_Q$ such that $H(0)=H_c$. 
Here, and in the following, we use single spin-flip dynamics and measure time in Monte Carlo (MC) steps per spin. 
Let the instantaneous relaxation time of the system be $\xi_t(t)$.
Initially, $\xi_t(t)$ is short and the evolution is adiabatic; the magnetization is thus close to its equilibrium value at $H(t)$.
However, the relaxation time diverges near the critical point, as does its rate of change. 
When its rate of change becomes larger than that of the system parameter (of order one for linear sweeps), the system can no longer stay in equilibrium: \mbox{$(\mathrm{d} \xi_t / \mathrm{d}t)|_{t=t_\mathrm{KZ}} \sim 1$}. 
This identifies the KZ time, $t_{\mathrm{KZ}}$. 
Crudely speaking, for \mbox{$t \in [-t_{\text{KZ}}, t_{\text{KZ}}]$}, the system is out of equilibrium, the dynamics are slow, and the magnetization response lags behind the field.
For $t \gg t_{\mathrm{KZ}}$, the evolution is once again adiabatic.
Over a full (forwards-backwards) sweep cycle, one thereby obtains a hysteresis loop in the magnetization $m$ as a function of $H$.
Examples of such hysteresis loops for different ramp times $\tau_Q$ are shown in Fig.~\ref{fig:bubble_field}.
The faster the sweep (the smaller $\tau_Q$), the larger the hysteresis loop.

For slow enough ramps 
($\tau_Q \gg 1$), 
generalized KZ scaling relations predict that the nonequilibrium contribution to the magnetization assumes a universal form~\cite{chandran_kibble-zurek_2012}: 
\begin{equation}
\langle  m (t) \rangle 
\sim 
\frac{1}{t_{\mathrm{KZ}}^{\Delta/z}}  \mathcal{G}\left(\frac{t}{t_{\mathrm{KZ}}}\right), \quad t_{\mathrm{KZ}} = \tau_Q^{\frac{\nu z}{\nu z +1}} 
, 
\label{eq:hysteresis_scaling}
\end{equation}
where $\Delta$ is the scaling dimension, and $\nu = (3 - \Delta)^{-1}$ the correlation length exponent, of the most relevant operator with projection on $m$ close to the critical point; $z$ is the dynamical exponent, and $\mathcal{G}$ is a universal scaling function. 
 
\begin{figure}[t!]
\includegraphics[width=1.0\linewidth]{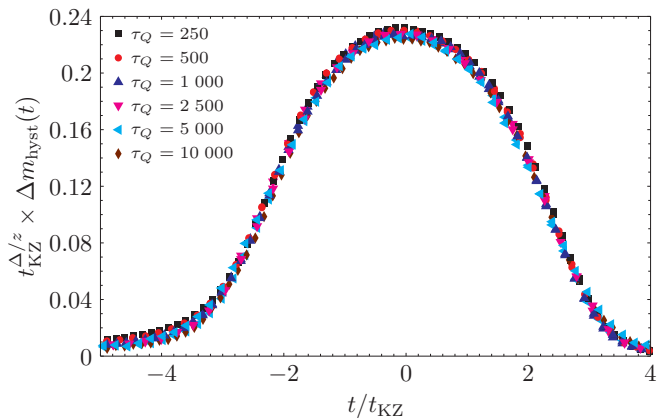}
\caption{\label{fig:kz_collapse}(Color online) Kibble-Zurek scaling collapse of the hysteresis loop size $\Delta m_\mathrm{hyst}(t)$,
using the critical exponents $\nu$ and $z$ as fitting parameters.
The collapse gives \mbox{$\nu= \nu_\sigma=0.42 \pm 0.01$} and \mbox{$z=1.85 \pm 0.05$}, consistent with the Ising values. Both the $L=8$ and 10 data are plotted.
}
\end{figure}

The size of the hysteresis loop---namely, the difference between forwards and backwards curves, $\Delta m_\mathrm{hyst}(t)$---can be calculated as a function of time to the critical point $t$ (inset of Fig.~\ref{fig:bubble_field}). 
Figure~\ref{fig:kz_collapse} shows the KZ scaling collapse of $\Delta m_\mathrm{hyst}(t)$ according to Eq.~\eqref{eq:hysteresis_scaling}, using the exponents $\nu$ and $z$ as fitting parameters. 
We find $\nu = \nu_\sigma = 0.42 \pm 0.01$ and $z= 1.85 \pm 0.05$, again consistent with the values for the 3D Ising model: $z \simeq 2.0$ for Metropolis single spin-flip dynamics~\cite{wansleben_monte_1991}; and (in nonzero field) the most relevant operator is the magnetization, with correlation length exponent $\nu_\sigma \simeq 0.40$~\cite{pelissetto_critical_2002}. 
Similar results are obtained for linear sweeps in generic directions in the $(T,H)$ plane.
We also checked robustness to small variations in the choice of the location of the critical point. 

In order to observe universal KZ scaling in numerical simulations of finite-size systems, one has to be in the appropriate speed regime~\cite{liu_dynamic_2014}. 
The system can generally be characterized by three length scales: a lattice scale $a$, the KZ length scale $l_{\textrm{KZ}} \sim t_{\textrm{KZ}}^{1/z}$, and the system size $L$.
Universal KZ physics characteristic of the thermodynamic limit appears when \mbox{$a \ll l_{\textrm{KZ}} \ll L$}.
Too slow sweeps lead to \mbox{$ l_{\textrm{KZ}}\gtrsim L$} and adiabatic evolution at finite size.
For the larger systems considered here ($L \geq 8$), the universal scaling behavior is clearly visible and spans approximately two decades in $\tau_Q$ (see Appendix \ref{app:ramp_speed}). 

It was recently demonstrated~\cite{kaiser_ac_2014} that spin ice systems, as magnetic Coulomb liquids, exhibit a (transient) second Wien effect, whereby the monopole density increases in response to an applied field. 
While this is in principle relevant to the dynamics discussed above, the effect is appreciable only at low monopole densities, which is not the case in our work. 

Finally, it is interesting to note that, even during our fastest sweeps, the spins located between adjacent kagome planes remain essentially fully polarized so that the monopole density $\rho$ follows extremely closely the magnetization $m$ of the system~\cite{mostame_tunable_2014} according to the formula \mbox{$m(\rho) \propto (5 \rho +10 ) /3$} (in units of $\mu_\mathrm{B}$ per spin). 
The magnetization is therefore an excellent proxy for the monopole density \footnote{We note that the defect (monopole) density in spin ice is not in a simple way related to the defect density a la Kibble-Zurek (which would be the defect density in the Ising order parameter). The monopole density is instead naturally related to the magnetisation, which is why it is possible for the former to be proportional to the latter close to the critical point, leading to the same scaling relations.}. 

\section{Outlook and experiments}

We have presented a study of the universal equilibrium and out-of-equilibrium properties of an emergent liquid-gas critical point in a three-dimensional topological magnet---spin ice in a [111] field. 
This holds the promise of experimental verification. 
One ingredient of practical importance is an unusually slow microscopic time scale due to the large energy barriers of the single-ion crystal-field environment, combined with comparably small transverse fields \cite{tomasello_single-ion_2015}. 
Spins appear to flip at a characteristic rate of approximately $1$~kHz~\cite{snyder_low-temperature_2004,jaubert_signature_2009}. 
This is to be contrasted with typical magnetic materials where microscopic dynamics occur on time scales of the order of nanoseconds or even picoseconds. 

Moreover, single spin-flip Monte Carlo dynamics have proved to capture reasonably well real dynamics in spin ice materials~\cite{jaubert_signature_2009, jaubert_magnetic_2011}. 
Combining these two observations to translate our results into experimentally relevant terms, we find that KZ scaling in spin ice may be accessible by sweeping fields in the range of $0.5$--$1$~T, at temperatures of the order of $0.6$~K, at rates from around $0.7$ to $0.03$~T/s, while measuring magnetization with an accuracy of about $1~\%$ or $\lesssim 0.1 \, \mu_\mathrm{B}$ per spin. 
This is eminently accessible and indeed comparable to achievements in earlier field sweep measurements~\cite{slobinsky_unconventional_2010}.

Spin ice may thus be the ideal experimental setup to observe dynamical Kibble-Zurek scaling in the context of a three dimensional topological magnet with a liquid-gas critical point driven by emergent fractionalized monopole excitations. 

\begin{acknowledgments}
	This work was supported in part by Engineering and Physical Sciences Research Council (EPSRC) Grant No.~EP/G049394/1 (C.C.), the Helmholtz Virtual Institute ``New States of Matter and Their Excitations,'' and the EPSRC NetworkPlus on ``Emergence and Physics far from Equilibrium.'' 
Research at Perimeter Institute is supported by the Government of Canada through Industry Canada and by the Province of Ontario through the Ministry of Economic Development and Innovation. 
The calculations were performed using the Darwin Supercomputer of the University of Cambridge High Performance Computing Service (\href{http://www.hpc.cam.ac.uk/}{http://www.hpc.cam.ac.uk/}) and the ARCHER UK National Supercomputing Service (\href{http://www.archer.ac.uk/}{http://www.archer.ac.uk/}, for which access was provided by the ARCHER Driving Test scheme).
The authors are grateful to A.~Sandvik for useful discussions and to S.~L.~Sondhi for advice and collaboration on several pieces of related work. J.O.H. is grateful to the EPSRC for funding, and to Michael Rutter for computing support.
Statement of compliance with EPSRC policy framework on research data: This publication reports theoretical work that does not require supporting research data. 
\end{acknowledgments}
%
%

%
%
\appendix

\section{Technical details of the Monte Carlo simulations}
\label{app:mc}

Our Monte Carlo simulations of spin ice use a conventional cubic unit cell containing 16 spins. 
A system consists of $L \times L \times L$ unit cells, so $N=16 L^3$ spins in total, with periodic boundary conditions. 
The Hamiltonian of the system is given by Eq.~\eqref{eq:hamiltonian} in the main text and we treat the long-range dipolar interactions using Ewald summation~\cite{leeuw_simulation_1980, melko_monte_2004}. 
We use the experimentally-determined parameters relevant for Dy$_2$Ti$_2$O$_7$, namely $J = -1.24$~K, and $D=1.41$~K from \onlinecite{den_hertog_dipolar_2000}, and we implement Metropolis single spin-flip updates. 

In our out-of-equilibrium simulations, we assume that Monte Carlo steps represent the actual time evolution of the system, namely, that the system has a well-defined single spin-flip time scale. 
This has been argued to correctly capture the dynamics in experiments at the temperatures relevant to this work~\cite{jaubert_signature_2009}.

\section{Location of the critical point}
\label{app:binder}

\begin{figure}[t!]
\includegraphics[width=1.0\linewidth]{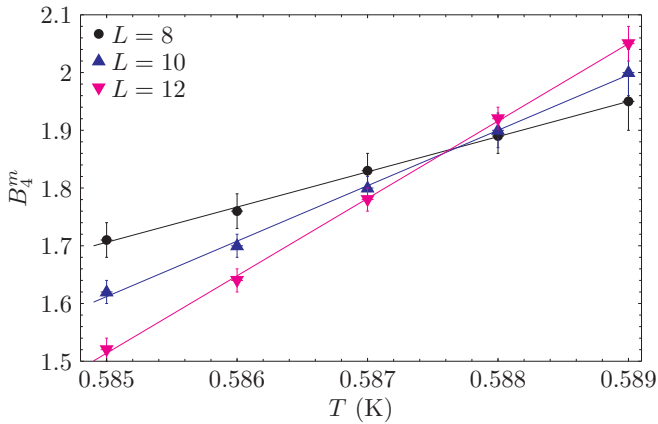}
\caption{\label{fig:binder_crossings}(Color online) Crossing of the fourth-order (Binder) cumulant of the magnetization, $B_4^m(T)$, for different system sizes $L$. 
The behavior is indicative of a continuous transition at \mbox{$T_c = 0.5875 \pm 0.0005$~K}. 
}
\end{figure}

As discussed in the main text, we analyze the fourth-order (Binder) cumulant~\cite{binder_finite_1981} to determine the location of the critical point. This cumulant is defined for any operator $X$ as 
\begin{equation}
B_4^X \equiv \frac{\langle ( \Delta X)^4 \rangle }{\langle(\Delta X)^2 \rangle^2} 
, 
\end{equation}
where $\Delta X \equiv X - \langle X \rangle$. 
The Binder cumulant of the magnetization $B_4^m$, at constant temperature $T$, shows a minimum at some value of the field which is the finite-size value for the transition field at that temperature and system size $L$.
Fig.~\ref{fig:binder_crossings} shows the minimum values of $B_4^m$ as a function of the temperature $T$ for different system sizes $L$ up to $L=12$ (27 648 spins).
The crossing for different system sizes is indicative of a continuous transition and the temperature at which it occurs gives an estimate for the critical temperature $T_c$. 
We find the crossing point of the two largest pairs of system sizes to be the same within errors (suggesting that finite-size effects for these system sizes are not the dominant source of error), and equal to $T_c=0.5875 \pm 0.0005$.
The estimate for the critical field $H_c$ is the transition field at $T_c$ as indicated by either the susceptibility maximum or Binder cumulant minimum. 
Our estimate for the location of the critical point is thus $(T_c,H_c) = (0.5875 \pm 0.0005 \, \mathrm{K}, 0.86295 \pm 0.00005 \, \mathrm{T})$.
We remark that the value of the Binder cumulant close to the critical point, $B_4^m=1.86 \pm 0.02$, differs from the infinite-volume Ising value exactly at criticality, $B_4^m \simeq 1.60$. 
Deviations in the value of the Binder cumulant have been reported in the literature due to details of the lattice structure and interactions~\cite{Schulte2005,Fenz2007}. 

We note that the magnetization $m$ used to obtain the Binder cumulant is not the critical Ising magnetizationlike operator $\tilde{m}$ due to field mixing (discussed in the main text).
However, identifying $\tilde{m}$ requires the values of $T_c$ and $H_c$, which in turn would in principle need the Binder cumulant of $\tilde{m}$.
Our analysis is justified because fluctuations of $\tilde{m}$ dominate close to the critical point, so $m$ or any quantity containing some $\tilde{m}$ can be used. 
For consistency, we repeated the Binder cumulant analysis using $\tilde{m}$ after obtaining the mixing parameters and checked that both the location of the critical point, and the value of the Binder cumulant at criticality, remain unchanged within error bars (not shown). 

\section{Further analysis of the ramp speed regimes}
\label{app:ramp_speed}

As discussed in the main text, in order to observe the universal out-of-equilibrium behavior characteristic of the thermodynamic limit in our sweeps, the correlation length when the system falls out of equilibrium, $l_{\mathrm{KZ}}$, should be much larger than the lattice constant $a$ (so that the dynamics are universal) but significantly smaller than the system size $L$ (so that the evolution is not adiabatic)~\cite{de_grandi_quench_2010,chandran_kibble-zurek_2012,huang_kibble-zurek_2014,huang_scaling_2015}. 

To investigate the different regimes systematically, we measure the total nonequilibrium contribution to the magnetization over a whole forwards-backwards sweep cycle, i.e., the integrated hysteresis loop area. 
Using the finite-time scaling relation given by Eq.~\eqref{eq:hysteresis_scaling} in the main text, the total area of the hysteresis loop scales with ramp time $\tau_Q$ as
\begin{equation}
\int \Delta m_\text{hyst}(t) \, dt \sim \tau_Q^{\left(1-\frac{\Delta}{z}\right) \left(\frac{\nu z }{\nu z +1} \right)}.
\end{equation} 

\begin{figure}[b!]
\includegraphics[width=1.0\linewidth]{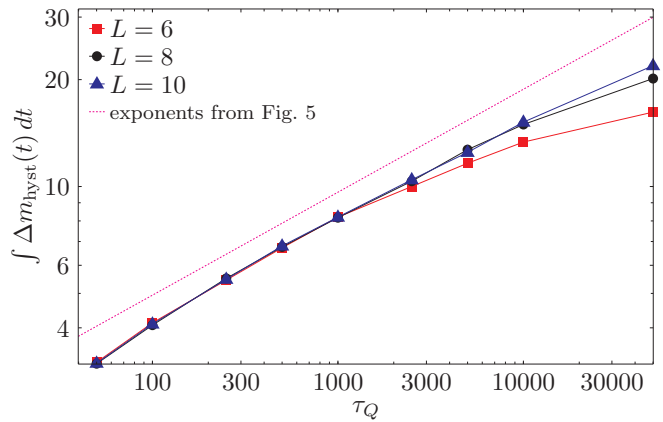}
\caption{\label{fig:area_vs_ramp_time}(Color online) Hysteresis loop area as a function of ramp time $\tau_Q$. 
Deviations from power-law scaling are seen at large $\tau_Q \sim L^z$ due to the time-evolution becoming adiabatic, and small $\tau_Q$ due to lattice-scale effects independent of $L$.
For intermediate $\tau_Q$ there is a universal scaling regime (pink dotted line with exponents from Fig.~\ref{fig:kz_collapse} in the main text).
}
\end{figure}

Fig.~\ref{fig:area_vs_ramp_time} shows the total hysteresis loop area as a function of ramp time $\tau_Q$, for different system sizes $L$.
Departures from power-law scaling can be seen at small and large $\tau_Q$.
The departure at large $\tau_Q$ is because the evolution is adiabatic when $l_\mathrm{KZ}$ is comparable to the system size.
Indeed, this departure occurs at smaller values of $\tau_Q$ for smaller $L$, as predicted by $l_{\mathrm{KZ}} \sim L$.
The departure at small $\tau_Q$ is because $l_{\mathrm{KZ}} \sim a$ and the evolution is nonuniversal.
Consistently, this departure is independent of system size. 
The nonuniversality in this fast sweep speed regime also manifests itself in an appreciable direction dependence in the nonequilibrium contribution to the magnetization (not shown). 
The universal scaling window is in the intermediate $\tau_Q$ regime. This window is appreciable (approximately two decades) for the largest system sizes we can access. The pink dotted line in Fig.~\ref{fig:area_vs_ramp_time} corresponds to scaling with the critical exponents from Fig.~\ref{fig:kz_collapse} in the main text.


\begin{thebibliography}{45}%
\makeatletter
\providecommand \@ifxundefined [1]{%
 \@ifx{#1\undefined}
}%
\providecommand \@ifnum [1]{%
 \ifnum #1\expandafter \@firstoftwo
 \else \expandafter \@secondoftwo
 \fi
}%
\providecommand \@ifx [1]{%
 \ifx #1\expandafter \@firstoftwo
 \else \expandafter \@secondoftwo
 \fi
}%
\providecommand \natexlab [1]{#1}%
\providecommand \enquote  [1]{``#1''}%
\providecommand \bibnamefont  [1]{#1}%
\providecommand \bibfnamefont [1]{#1}%
\providecommand \citenamefont [1]{#1}%
\providecommand \href@noop [0]{\@secondoftwo}%
\providecommand \href [0]{\begingroup \@sanitize@url \@href}%
\providecommand \@href[1]{\@@startlink{#1}\@@href}%
\providecommand \@@href[1]{\endgroup#1\@@endlink}%
\providecommand \@sanitize@url [0]{\catcode `\\12\catcode `\$12\catcode
  `\&12\catcode `\#12\catcode `\^12\catcode `\_12\catcode `\%12\relax}%
\providecommand \@@startlink[1]{}%
\providecommand \@@endlink[0]{}%
\providecommand \url  [0]{\begingroup\@sanitize@url \@url }%
\providecommand \@url [1]{\endgroup\@href {#1}{\urlprefix }}%
\providecommand \urlprefix  [0]{URL }%
\providecommand \Eprint [0]{\href }%
\@ifxundefined \urlstyle {%
  \providecommand \doi  [0]{\begingroup \@sanitize@url \@doi}%
  \providecommand \@doi [1]{\endgroup \@@startlink {\doibase
  #1}doi:\discretionary {}{}{}#1\@@endlink }%
}{%
  \providecommand \doi  [0]{doi:\discretionary{}{}{}\begingroup
  \urlstyle{rm}\Url }%
}%
\providecommand \doibase [0]{http://dx.doi.org/}%
\providecommand \Doi [0]{\begingroup \@sanitize@url \@Doi }%
\providecommand \@Doi  [1]{\endgroup\@@startlink{\doibase#1}\@@Doi}%
\providecommand \@@Doi [1]{#1\@@endlink}%
\providecommand \selectlanguage [0]{\@gobble}%
\providecommand \bibinfo  [0]{\@secondoftwo}%
\providecommand \bibfield  [0]{\@secondoftwo}%
\providecommand \translation [1]{[#1]}%
\providecommand \BibitemOpen [0]{}%
\providecommand \bibitemStop [0]{}%
\providecommand \bibitemNoStop [0]{.\EOS\space}%
\providecommand \EOS [0]{\spacefactor3000\relax}%
\providecommand \BibitemShut  [1]{\csname bibitem#1\endcsname}%
\bibitem [{\citenamefont {Goldenfeld}(1992)}]{goldenfeld_lectures_1992}%
  \BibitemOpen
  \bibfield  {author} {\bibinfo {author} {\bibfnamefont {N.}~\bibnamefont
  {Goldenfeld}},\ }\href@noop {} {\emph {\bibinfo {title} {Lectures on Phase
  Transitions and the Renormalization Group}}}\ (\bibinfo  {publisher}
  {Addison-Wesley, Reading},\ \bibinfo {year} {1992})\BibitemShut {NoStop}%
\bibitem [{\citenamefont {Hohenberg}\ and\ \citenamefont
  {Halperin}(1977)}]{Hohenberg:1977aa}%
  \BibitemOpen
  \bibfield  {author} {\bibinfo {author} {\bibfnamefont {P.~C.}\ \bibnamefont
  {Hohenberg}}\ and\ \bibinfo {author} {\bibfnamefont {B.~I.}\ \bibnamefont
  {Halperin}},\ }\Doi {10.1103/RevModPhys.49.435} {\bibfield  {journal}
  {\bibinfo  {journal} {Rev.~Mod.~Phys.},\ }\textbf {\bibinfo {volume} {49}},\
  \bibinfo {pages} {435} (\bibinfo {year} {1977})}\BibitemShut {NoStop}%
\bibitem [{\citenamefont {Dziarmaga}(2010)}]{Dziarmaga:2010qf}%
  \BibitemOpen
  \bibfield  {author} {\bibinfo {author} {\bibfnamefont {J.}~\bibnamefont
  {Dziarmaga}},\ }\Doi {10.1080/00018732.2010.514702} {\bibfield  {journal}
  {\bibinfo  {journal} {Adv.~Phys.},\ }\textbf {\bibinfo {volume} {59}},\
  \bibinfo {pages} {1063} (\bibinfo {year} {2010})}\BibitemShut {NoStop}%
\bibitem [{\citenamefont {Polkovnikov}\ \emph {et~al.}(2011)\citenamefont
  {Polkovnikov}, \citenamefont {Sengupta}, \citenamefont {Silva},\ and\
  \citenamefont {Vengalattore}}]{polkovnikov_colloquium:_2011}%
  \BibitemOpen
  \bibfield  {author} {\bibinfo {author} {\bibfnamefont {A.}~\bibnamefont
  {Polkovnikov}}, \bibinfo {author} {\bibfnamefont {K.}~\bibnamefont
  {Sengupta}}, \bibinfo {author} {\bibfnamefont {A.}~\bibnamefont {Silva}}, \
  and\ \bibinfo {author} {\bibfnamefont {M.}~\bibnamefont {Vengalattore}},\
  }\Doi {10.1103/RevModPhys.83.863} {\bibfield  {journal} {\bibinfo  {journal}
  {Rev.~Mod.~Phys.},\ }\textbf {\bibinfo {volume} {83}},\ \bibinfo {pages}
  {863} (\bibinfo {year} {2011})}\BibitemShut {NoStop}%
\bibitem [{\citenamefont {del Campo}\ and\ \citenamefont
  {Zurek}(2014)}]{del_campo_universality_2014}%
  \BibitemOpen
  \bibfield  {author} {\bibinfo {author} {\bibfnamefont {A.}~\bibnamefont {del
  Campo}}\ and\ \bibinfo {author} {\bibfnamefont {W.~H.}\ \bibnamefont
  {Zurek}},\ }\Doi {10.1142/S0217751X1430018X} {\bibfield  {journal} {\bibinfo
  {journal} {Int.~J.~Mod.~Phys.~A},\ }\textbf {\bibinfo {volume} {29}},\
  \bibinfo {pages} {1430018} (\bibinfo {year} {2014})}\BibitemShut {NoStop}%
\bibitem [{\citenamefont {Wen}(2004)}]{Wen_book_2004}%
  \BibitemOpen
  \bibfield  {author} {\bibinfo {author} {\bibfnamefont {X.-G.}\ \bibnamefont
  {Wen}},\ }\href@noop {} {\emph {\bibinfo {title} {Quantum Field Theory of
  Many-Body Systems}}}\ (\bibinfo  {publisher} {OUP, Oxford UK},\ \bibinfo
  {year} {2004})\BibitemShut {NoStop}%
\bibitem [{\citenamefont {Senthil}\ \emph {et~al.}(2004)\citenamefont
  {Senthil} \emph {et~al.}}]{senthil_deconfined_2004}%
  \BibitemOpen
  \bibfield  {author} {\bibinfo {author} {\bibfnamefont {T.}~\bibnamefont
  {Senthil}} \emph {et~al.},\ }\Doi {10.1126/science.1091806}
  {\bibfield  {journal} {\bibinfo  {journal} {Science},\ }\textbf
  {\bibinfo {volume} {303}},\ \bibinfo {pages} {1490} (\bibinfo {year}
  {2004})}\BibitemShut {NoStop}%
\bibitem [{\citenamefont {Moessner}\ \emph {et~al.}(2001)\citenamefont
  {Moessner} \emph {et~al.}}]{moessner_sondhi_fradkin_2001}%
  \BibitemOpen
  \bibfield  {author} {\bibinfo {author} {\bibfnamefont {R.}~\bibnamefont
  {Moessner}}, \bibinfo {author} {\bibfnamefont {S.~L.}\ \bibnamefont
  {Sondhi}}\ and\ \bibinfo {author} {\bibfnamefont {E.}\ \bibnamefont
  {Fradkin}},\ }\Doi {10.1103/PhysRevB.65.024504}
  {\bibfield  {journal} {\bibinfo  {journal} {Phys.~Rev.~B},\ }\textbf
  {\bibinfo {volume} {65}},\ \bibinfo {pages} {024504} (\bibinfo {year}
  {2001})}\BibitemShut {NoStop}%
\bibitem [{Note1()}]{Note1}%
  \BibitemOpen
  \bibinfo {note} {For a collection of current topics, see e.g., \protect \href
  {http://www.pks.mpg.de/\textasciitilde
  tomaeq14/}{http://www.pks.mpg.de/\textasciitilde tomaeq14/}}\BibitemShut
  {NoStop}%
\bibitem [{\citenamefont {Bramwell}\ and\ \citenamefont
  {Gingras}(2001)}]{bramwell_spin_2001}%
  \BibitemOpen
  \bibfield  {author} {\bibinfo {author} {\bibfnamefont {S.~T.}\ \bibnamefont
  {Bramwell}}\ and\ \bibinfo {author} {\bibfnamefont {M.~J.~P.}\ \bibnamefont
  {Gingras}},\ }\Doi {10.1126/science.1064761} {\bibfield  {journal} {\bibinfo
  {journal} {Science},\ }\textbf {\bibinfo {volume} {294}},\ \bibinfo {pages}
  {1495} (\bibinfo {year} {2001})}\BibitemShut {NoStop}%
\bibitem [{\citenamefont {Castelnovo}\ \emph {et~al.}(2012)\citenamefont
  {Castelnovo}, \citenamefont {Moessner},\ and\ \citenamefont
  {Sondhi}}]{castelnovo_spin_2012}%
  \BibitemOpen
  \bibfield  {author} {\bibinfo {author} {\bibfnamefont {C.}~\bibnamefont
  {Castelnovo}}, \bibinfo {author} {\bibfnamefont {R.}~\bibnamefont
  {Moessner}}, \ and\ \bibinfo {author} {\bibfnamefont {S.}~\bibnamefont
  {Sondhi}},\ }\Doi {10.1146/annurev-conmatphys-020911-125058} {\bibfield
  {journal} {\bibinfo  {journal} {Ann.~Rev.~Cond.~Mat.~Phys.},\ }\textbf
  {\bibinfo {volume} {3}},\ \bibinfo {pages} {35} (\bibinfo {year}
  {2012})}\BibitemShut {NoStop}%
\bibitem [{\citenamefont {Castelnovo}\ \emph {et~al.}(2011)\citenamefont
  {Castelnovo}, \citenamefont {Moessner},\ and\ \citenamefont
  {Sondhi}}]{castelnovo_DH_2011}%
  \BibitemOpen
  \bibfield  {author} {\bibinfo {author} {\bibfnamefont {C.}~\bibnamefont
  {Castelnovo}}, \bibinfo {author} {\bibfnamefont {R.}~\bibnamefont
  {Moessner}}, \ and\ \bibinfo {author} {\bibfnamefont {S.~L.}~\bibnamefont
  {Sondhi}},\ }\Doi {10.1103/PhysRevB.84.144435} {\bibfield
  {journal} {\bibinfo  {journal} {Phys.~Rev.~B},\ }\textbf
  {\bibinfo {volume} {84}},\ \bibinfo {pages} {144435} (\bibinfo {year}
  {2011})}\BibitemShut {NoStop}%
\bibitem [{\citenamefont {Castelnovo}\ \emph {et~al.}(2010)\citenamefont
  {Castelnovo}, \citenamefont {Moessner},\ and\ \citenamefont
  {Sondhi}}]{castelnovo_thquench_2010}%
  \BibitemOpen
  \bibfield  {author} {\bibinfo {author} {\bibfnamefont {C.}~\bibnamefont
  {Castelnovo}}, \bibinfo {author} {\bibfnamefont {R.}~\bibnamefont
  {Moessner}}, \ and\ \bibinfo {author} {\bibfnamefont {S.~L.}~\bibnamefont
  {Sondhi}},\ }\Doi {10.1103/PhysRevLett.104.107201} {\bibfield
  {journal} {\bibinfo  {journal} {Phys.~Rev.~Lett.},\ }\textbf
  {\bibinfo {volume} {104}},\ \bibinfo {pages} {107201} (\bibinfo {year}
  {2010})}\BibitemShut {NoStop}%
\bibitem [{\citenamefont {Kaiser}\ \emph {et~al.}(2014)\citenamefont {Kaiser},
  \citenamefont {Bramwell}, \citenamefont {Holdsworth},\ and\ \citenamefont
  {Moessner}}]{kaiser_ac_2014}%
  \BibitemOpen
  \bibfield  {author} {\bibinfo {author} {\bibfnamefont {V.}~\bibnamefont
  {Kaiser}}, \bibinfo {author} {\bibfnamefont {S.~T.}\ \bibnamefont
  {Bramwell}}, \bibinfo {author} {\bibfnamefont {P.~C.~W.}\ \bibnamefont
  {Holdsworth}}, \ and\ \bibinfo {author} {\bibfnamefont {R.}~\bibnamefont
{Moessner}},\ } \Doi {10.1103/PhysRevLett.115.037201} {\bibfield {journal} {\bibinfo {journal} {Phys.~Rev.~Lett.},\ }\textbf {\bibinfo {volume} {115}},\ \bibinfo {pages} {037201} (\bibinfo {year} {2015})} \BibitemShut {NoStop}%
\bibitem [{\citenamefont {Sakakibara}\ \emph {et~al.}(2003)\citenamefont
  {Sakakibara} \emph {et~al.}}]{sakakibara_observation_2003}%
  \BibitemOpen
  \bibfield  {author} {\bibinfo {author} {\bibfnamefont {T.}~\bibnamefont
	{Sakakibara}} \bibinfo {author} {\bibnamefont {T.}~\bibnamefont {Tayama}}, \bibinfo {author} {\bibnamefont {Z.}~\bibnamefont {Hiroi}}, \bibinfo {author} {\bibnamefont {K.}~\bibnamefont {Matsuhira}}, \ and\ \bibinfo{author}{\bibnamefont {S.}~\bibnamefont {Takagi}},\ }\Doi {10.1103/PhysRevLett.90.207205}
  {\bibfield  {journal} {\bibinfo  {journal} {Phys.~Rev.~Lett.},\ }\textbf
  {\bibinfo {volume} {90}},\ \bibinfo {pages} {207205} (\bibinfo {year}
  {2003})}\BibitemShut {NoStop}%
\bibitem [{\citenamefont {Castelnovo}\ \emph {et~al.}(2008)\citenamefont
  {Castelnovo}, \citenamefont {Moessner},\ and\ \citenamefont
  {Sondhi}}]{castelnovo_magnetic_2008}%
  \BibitemOpen
  \bibfield  {author} {\bibinfo {author} {\bibfnamefont {C.}~\bibnamefont
  {Castelnovo}}, \bibinfo {author} {\bibfnamefont {R.}~\bibnamefont
  {Moessner}}, \ and\ \bibinfo {author} {\bibfnamefont {S.~L.}\ \bibnamefont
  {Sondhi}},\ }\Doi {10.1038/nature06433} {\bibfield  {journal} {\bibinfo
  {journal} {Nature},\ }\textbf {\bibinfo {volume} {451}},\ \bibinfo {pages}
  {42} (\bibinfo {year} {2008})}\BibitemShut {NoStop}%
\bibitem [{\citenamefont {Fisher}(1994)}]{fisher_story_1994}%
  \BibitemOpen
  \bibfield  {author} {\bibinfo {author} {\bibfnamefont {M.~E.}\ \bibnamefont
  {Fisher}},\ }\Doi {10.1007/BF02186278} {\bibfield  {journal} {\bibinfo
  {journal} {J.~Stat.~Phys.},\ }\textbf {\bibinfo {volume} {75}},\ \bibinfo
  {pages} {1} (\bibinfo {year} {1994})}\BibitemShut {NoStop}%
\bibitem [{\citenamefont {Wilding}(1997)}]{wilding_simulation_1997}%
  \BibitemOpen
  \bibfield  {author} {\bibinfo {author} {\bibfnamefont {N.~B.}\ \bibnamefont
  {Wilding}},\ }\Doi {10.1088/0953-8984/9/3/002} {\bibfield  {journal}
  {\bibinfo  {journal} {J.~Phys.: Cond.~Mat.},\ }\textbf {\bibinfo {volume}
  {9}},\ \bibinfo {pages} {585} (\bibinfo {year} {1997})}\BibitemShut {NoStop}%
\bibitem [{\citenamefont {Luijten}\ \emph {et~al.}(2002)\citenamefont
  {Luijten}, \citenamefont {Fisher},\ and\ \citenamefont
  {Panagiotopoulos}}]{luijten_universality_2002}%
  \BibitemOpen
  \bibfield  {author} {\bibinfo {author} {\bibfnamefont {E.}~\bibnamefont
  {Luijten}}, \bibinfo {author} {\bibfnamefont {M.~E.}\ \bibnamefont {Fisher}},
  \ and\ \bibinfo {author} {\bibfnamefont {A.~Z.}\ \bibnamefont
  {Panagiotopoulos}},\ }\Doi {10.1103/PhysRevLett.88.185701} {\bibfield
  {journal} {\bibinfo  {journal} {Phys.~Rev.~Lett.},\ }\textbf {\bibinfo
  {volume} {88}},\ \bibinfo {pages} {185701} (\bibinfo {year}
  {2002})}\BibitemShut {NoStop}%
\bibitem [{\citenamefont {Kibble}(1976)}]{kibble_topology_1976}%
  \BibitemOpen
  \bibfield  {author} {\bibinfo {author} {\bibfnamefont {T.~W.~B.}\
  \bibnamefont {Kibble}},\ }\Doi {10.1088/0305-4470/9/8/029} {\bibfield
  {journal} {\bibinfo  {journal} {J.~Phys.~A},\ }\textbf {\bibinfo {volume}
  {9}},\ \bibinfo {pages} {1387} (\bibinfo {year} {1976})}\BibitemShut
  {NoStop}%
\bibitem [{\citenamefont {Zurek}(1985)}]{zurek_cosmological_1985}%
  \BibitemOpen
  \bibfield  {author} {\bibinfo {author} {\bibfnamefont {W.~H.}\ \bibnamefont
  {Zurek}},\ }\Doi {10.1038/317505a0} {\bibfield  {journal} {\bibinfo
  {journal} {Nature},\ }\textbf {\bibinfo {volume} {317}},\ \bibinfo {pages}
  {505} (\bibinfo {year} {1985})}\BibitemShut {NoStop}%
\bibitem [{\citenamefont {Polkovnikov}(2005)}]{Polkovnikov2005}%
  \BibitemOpen
  \bibfield  {author} {\bibinfo {author} {\bibfnamefont {A.}~\bibnamefont
  {Polkovnikov}},\ }\Doi {10.1103/PhysRevB.72.161201} {\bibfield  {journal}
  {\bibinfo  {journal} {Phys. Rev. B},\ }\textbf {\bibinfo {volume} {72}},\
  \bibinfo {pages} {161201} (\bibinfo {year} {2005})}\BibitemShut {NoStop}%
\bibitem [{\citenamefont {Deng}\ \emph {et~al.}(2008)\citenamefont {Deng},
  \citenamefont {Ortiz},\ and\ \citenamefont {Viola}}]{Deng-S.:2008aa}%
  \BibitemOpen
  \bibfield  {author} {\bibinfo {author} {\bibfnamefont {S.}~\bibnamefont
  {Deng}}, \bibinfo {author} {\bibfnamefont {G.}~\bibnamefont {Ortiz}}, \ and\
  \bibinfo {author} {\bibfnamefont {L.}~\bibnamefont {Viola}},\ }\Doi
  {10.1209/0295-5075/84/67008} {\bibfield  {journal} {\bibinfo  {journal}
  {EPL},\ }\textbf {\bibinfo {volume} {84}},\ \bibinfo {pages} {67008}
  (\bibinfo {year} {2008})}\BibitemShut {NoStop}%
\bibitem [{\citenamefont {Biroli}\ \emph {et~al.}(2010)\citenamefont {Biroli},
  \citenamefont {Cugliandolo},\ and\ \citenamefont
  {Sicilia}}]{biroli_kibble-zurek_2010}%
  \BibitemOpen
  \bibfield  {author} {\bibinfo {author} {\bibfnamefont {G.}~\bibnamefont
  {Biroli}}, \bibinfo {author} {\bibfnamefont {L.~F.}\ \bibnamefont
  {Cugliandolo}}, \ and\ \bibinfo {author} {\bibfnamefont {A.}~\bibnamefont
  {Sicilia}},\ }\Doi {10.1103/PhysRevE.81.050101} {\bibfield  {journal}
  {\bibinfo  {journal} {Phys.~Rev.~E},\ }\textbf {\bibinfo {volume} {81}},\
  \bibinfo {pages} {050101} (\bibinfo {year} {2010})}\BibitemShut {NoStop}%
\bibitem [{\citenamefont {Chandran}\ \emph {et~al.}(2012)\citenamefont
	{Chandran}, \citenamefont {Erez}, \citenamefont {Gubser},\ and\ \citenamefont{Sondhi}}]{chandran_kibble-zurek_2012}%
  \BibitemOpen
  \bibfield  {author} {\bibinfo {author} {\bibnamefont {A.}~\bibnamefont
	{Chandran}}, \bibinfo {author} {\bibnamefont {A.}~\bibnamefont {Erez}}, \bibinfo {author} {\bibnamefont {S.~S.} \bibnamefont {Gubser}},\ and\ \bibinfo {author} {\bibnamefont {S.~L.} \bibnamefont {Sondhi}},\ }\Doi {10.1103/PhysRevB.86.064304} {\bibfield
  {journal} {\bibinfo  {journal} {Phys.~Rev.~B},\ }\textbf {\bibinfo {volume}
  {86}},\ \bibinfo {pages} {064304} (\bibinfo {year} {2012})}\BibitemShut
  {NoStop}%
\bibitem [{\citenamefont {Slobinsky}\ \emph {et~al.}(2010)\citenamefont
	{Slobinsky}, \citenamefont {Castelnovo}, \citenamefont {Borzi}, \citenamefont {Gibbs}, \citenamefont {Mackenzie}, \citenamefont {Moessner},\ and\ \citenamefont {Grigera}}]{slobinsky_unconventional_2010}%
  \BibitemOpen
  \bibfield  {author} {\bibinfo {author} {\bibfnamefont {D.}~\bibnamefont
	{Slobinsky}}, \bibinfo {author}{\bibnamefont {C.}~\bibnamefont {Castelnovo}}, \bibinfo {author} {\bibnamefont {R.~A.}~\bibnamefont {Borzi}}, \bibinfo {author} {\bibnamefont {A.~S.}~\bibnamefont {Gibbs}}, \bibinfo {author}{\bibnamefont {A.~P.}\ \bibnamefont {Mackenzie}}, \bibinfo {author}{\bibnamefont {R.}~\bibnamefont {Moessner}},\ and\ \bibinfo {author}{\bibnamefont{S.~A.}~\bibnamefont {Grigera}},\ }\Doi {10.1103/PhysRevLett.105.267205}
  {\bibfield  {journal} {\bibinfo  {journal} {Phys.~Rev.~Lett.},\ }\textbf
  {\bibinfo {volume} {105}},\ \bibinfo {pages} {267205} (\bibinfo {year}
  {2010})}\BibitemShut {NoStop}%
\bibitem [{\citenamefont {Snyder}\ \emph {et~al.}(2004)\citenamefont {Snyder}
\citenamefont {Ueland}, \citenamefont {Slusky}, \citenamefont {Karunadasa}, \citenamefont {Cava},\ and\ \citenamefont {Schiffer}}]{snyder_low-temperature_2004}%
  \BibitemOpen
  \bibfield  {author} {\bibinfo {author} {\bibfnamefont {J.}~\bibnamefont
	{Snyder}} \bibinfo {author}{\bibnamefont {B.~G.~}\bibnamefont {Ueland}}, \bibinfo {author}{\bibnamefont {J.~S.~}\bibnamefont {Slusky}}, \bibinfo {author}{\bibnamefont {H.~}\bibnamefont {Karunadasa}}, \bibinfo {author}{\bibnamefont {R.~J.~}\bibnamefont {Cava}},\ and\ \bibinfo {author}{\bibnamefont {P.~}\bibnamefont {Schiffer}},\ }\Doi {10.1103/PhysRevB.69.064414} {\bibfield
  {journal} {\bibinfo  {journal} {Phys.~Rev.~B},\ }\textbf {\bibinfo {volume}
  {69}},\ \bibinfo {pages} {064414} (\bibinfo {year} {2004})}\BibitemShut
  {NoStop}%
\bibitem [{\citenamefont {Jaubert}\ and\ \citenamefont
  {Holdsworth}(2009)}]{jaubert_signature_2009}%
  \BibitemOpen
  \bibfield  {author} {\bibinfo {author} {\bibfnamefont {L.~D.~C.}\
  \bibnamefont {Jaubert}}\ and\ \bibinfo {author} {\bibfnamefont {P.~C.~W.}\
  \bibnamefont {Holdsworth}},\ }\Doi {10.1038/nphys1227} {\bibfield  {journal}
  {\bibinfo  {journal} {Nature Physics},\ }\textbf {\bibinfo {volume} {5}},\
  \bibinfo {pages} {258} (\bibinfo {year} {2009})}\BibitemShut {NoStop}%
\bibitem [{\citenamefont {Schwinger}(1951)}]{Schwinger1951}%
  \BibitemOpen
  \bibfield  {author} {\bibinfo {author} {\bibfnamefont {J.}~\bibnamefont
  {Schwinger}},\ }\Doi {10.1103/PhysRev.82.664} {\bibfield  {journal} {\bibinfo
   {journal} {Phys. Rev.},\ }\textbf {\bibinfo {volume} {82}},\ \bibinfo
  {pages} {664} (\bibinfo {year} {1951})}\BibitemShut {NoStop}%
\bibitem [{\citenamefont {den Hertog}\ and\ \citenamefont
  {Gingras}(2000)}]{den_hertog_dipolar_2000}%
  \BibitemOpen
  \bibfield  {author} {\bibinfo {author} {\bibfnamefont {B.~C.}\ \bibnamefont
  {den Hertog}}\ and\ \bibinfo {author} {\bibfnamefont {M.~J.~P.}\ \bibnamefont
  {Gingras}},\ }\Doi {10.1103/PhysRevLett.84.3430} {\bibfield  {journal}
  {\bibinfo  {journal} {Phys.~Rev.~Lett.},\ }\textbf {\bibinfo {volume} {84}},\
  \bibinfo {pages} {3430} (\bibinfo {year} {2000})}\BibitemShut {NoStop}%
\bibitem [{\citenamefont {Rehr}\ and\ \citenamefont
  {Mermin}(1973)}]{rehr_revised_1973}%
  \BibitemOpen
  \bibfield  {author} {\bibinfo {author} {\bibfnamefont {J.~J.}\ \bibnamefont
  {Rehr}}\ and\ \bibinfo {author} {\bibfnamefont {N.~D.}\ \bibnamefont
  {Mermin}},\ }\Doi {10.1103/PhysRevA.8.472} {\bibfield  {journal} {\bibinfo
  {journal} {Phys.~Rev.~A},\ }\textbf {\bibinfo {volume} {8}},\ \bibinfo
  {pages} {472} (\bibinfo {year} {1973})}\BibitemShut {NoStop}%
\bibitem [{\citenamefont {Hiroi}\ \emph {et~al.}(2003)\citenamefont {Hiroi}
  \emph {et~al.}}]{hiroi_specific_2003}%
  \BibitemOpen
  \bibfield  {author} {\bibinfo {author} {\bibfnamefont {Z.}~\bibnamefont
  {Hiroi}} \emph {et~al.},\ }\Doi {10.1143/JPSJ.72.411} {\bibfield  {journal}
  {\bibinfo  {journal} {J.~Phys.~Soc.~Jpn},\ }\textbf {\bibinfo {volume}
  {72}},\ \bibinfo {pages} {411} (\bibinfo {year} {2003})}\BibitemShut
  {NoStop}%
\bibitem [{\citenamefont {Rummukainen}\ \emph {et~al.}(1998)\citenamefont
  {Rummukainen} \emph {et~al.}}]{rummukainen_universality_1998}%
  \BibitemOpen
  \bibfield  {author} {\bibinfo {author} {\bibfnamefont {K.}~\bibnamefont
  {Rummukainen}} \emph {et~al.},\ }\Doi {10.1016/S0550-3213(98)00494-5}
  {\bibfield  {journal} {\bibinfo  {journal} {Nuc.~Phys.~B},\ }\textbf
  {\bibinfo {volume} {532}},\ \bibinfo {pages} {283} (\bibinfo {year}
  {1998})}\BibitemShut {NoStop}%
\bibitem [{\citenamefont {Bruce}(1981)}]{bruce_probability_1981}%
  \BibitemOpen
  \bibfield  {author} {\bibinfo {author} {\bibfnamefont {A.~D.}\ \bibnamefont
  {Bruce}},\ }\Doi {10.1088/0022-3719/14/25/012} {\bibfield  {journal}
  {\bibinfo  {journal} {J.~Phys~C: Solid State Physics},\ }\textbf {\bibinfo
  {volume} {14}},\ \bibinfo {pages} {3667} (\bibinfo {year}
  {1981})}\BibitemShut {NoStop}%
\bibitem [{\citenamefont {Plascak}\ and\ \citenamefont
  {Martins}(2013)}]{plascak_probability_2013}%
  \BibitemOpen
  \bibfield  {author} {\bibinfo {author} {\bibfnamefont {J.~A.}\ \bibnamefont
  {Plascak}}\ and\ \bibinfo {author} {\bibfnamefont {P.~H.~L.}\ \bibnamefont
  {Martins}},\ }\Doi {10.1016/j.cpc.2012.09.014} {\bibfield  {journal}
  {\bibinfo  {journal} {Comp.~Phys.~Comms.},\ }\textbf {\bibinfo {volume}
  {184}},\ \bibinfo {pages} {259} (\bibinfo {year} {2013})}\BibitemShut
  {NoStop}%
\bibitem [{\citenamefont {Wilding}\ and\ \citenamefont
  {M\"uller}(1995)}]{wilding_liquidvapor_1995}%
  \BibitemOpen
  \bibfield  {author} {\bibinfo {author} {\bibfnamefont {N.~B.}\ \bibnamefont
  {Wilding}}\ and\ \bibinfo {author} {\bibfnamefont {M.}~\bibnamefont
  {M\"uller}},\ }\Doi {10.1063/1.468686} {\bibfield  {journal} {\bibinfo
  {journal} {J.~Chem.~Phys.},\ }\textbf {\bibinfo {volume} {102}},\ \bibinfo
  {pages} {2562} (\bibinfo {year} {1995})}\BibitemShut {NoStop}%
\bibitem [{\citenamefont {Karsch}\ and\ \citenamefont
  {Stickan}(2000)}]{karsch_three-dimensional_2000}%
  \BibitemOpen
  \bibfield  {author} {\bibinfo {author} {\bibfnamefont {F.}~\bibnamefont
  {Karsch}}\ and\ \bibinfo {author} {\bibfnamefont {S.}~\bibnamefont
  {Stickan}},\ }\Doi {10.1016/S0370-2693(00)00902-3} {\bibfield  {journal}
  {\bibinfo  {journal} {Phys.~Lett.~B},\ }\textbf {\bibinfo {volume} {488}},\
  \bibinfo {pages} {319} (\bibinfo {year} {2000})}\BibitemShut {NoStop}%
\bibitem [{\citenamefont {Liu}\ \emph {et~al.}(2014)\citenamefont {Liu},
  \citenamefont {Polkovnikov},\ and\ \citenamefont
  {Sandvik}}]{liu_dynamic_2014}%
  \BibitemOpen
  \bibfield  {author} {\bibinfo {author} {\bibfnamefont {C.-W.}\ \bibnamefont
  {Liu}}, \bibinfo {author} {\bibfnamefont {A.}~\bibnamefont {Polkovnikov}}, \
  and\ \bibinfo {author} {\bibfnamefont {A.~W.}\ \bibnamefont {Sandvik}},\
  }\Doi {10.1103/PhysRevB.89.054307} {\bibfield  {journal} {\bibinfo  {journal}
  {Phys.~Rev.~B},\ }\textbf {\bibinfo {volume} {89}},\ \bibinfo {pages}
  {054307} (\bibinfo {year} {2014})}\BibitemShut {NoStop}%
\bibitem [{\citenamefont {Wansleben}\ and\ \citenamefont
  {Landau}(1991)}]{wansleben_monte_1991}%
  \BibitemOpen
  \bibfield  {author} {\bibinfo {author} {\bibfnamefont {S.}~\bibnamefont
  {Wansleben}}\ and\ \bibinfo {author} {\bibfnamefont {D.~P.}\ \bibnamefont
  {Landau}},\ }\Doi {10.1103/PhysRevB.43.6006} {\bibfield  {journal} {\bibinfo
  {journal} {Phys.~Rev.~B},\ }\textbf {\bibinfo {volume} {43}},\ \bibinfo
  {pages} {6006} (\bibinfo {year} {1991})}\BibitemShut {NoStop}%
\bibitem [{\citenamefont {Pelissetto}\ and\ \citenamefont
  {Vicari}(2002)}]{pelissetto_critical_2002}%
  \BibitemOpen
  \bibfield  {author} {\bibinfo {author} {\bibfnamefont {A.}~\bibnamefont
  {Pelissetto}}\ and\ \bibinfo {author} {\bibfnamefont {E.}~\bibnamefont
  {Vicari}},\ }\Doi {10.1016/S0370-1573(02)00219-3} {\bibfield  {journal}
  {\bibinfo  {journal} {Phys.~Rep.},\ }\textbf {\bibinfo {volume} {368}},\
  \bibinfo {pages} {549} (\bibinfo {year} {2002})}\BibitemShut {NoStop}%
\bibitem [{\citenamefont {Mostame}\ \emph {et~al.}(2014)\citenamefont
  {Mostame}, \citenamefont {Castelnovo}, \citenamefont {Moessner},\ and\
  \citenamefont {Sondhi}}]{mostame_tunable_2014}%
  \BibitemOpen
  \bibfield  {author} {\bibinfo {author} {\bibfnamefont {S.}~\bibnamefont
  {Mostame}}, \bibinfo {author} {\bibfnamefont {C.}~\bibnamefont {Castelnovo}},
  \bibinfo {author} {\bibfnamefont {R.}~\bibnamefont {Moessner}}, \ and\
  \bibinfo {author} {\bibfnamefont {S.~L.}\ \bibnamefont {Sondhi}},\ }\Doi
  {10.1073/pnas.1317631111} {\bibfield  {journal} {\bibinfo  {journal}
  {Proc.~Nat.~Acad.~Sci.},\ }\textbf {\bibinfo {volume} {111}},\ \bibinfo
  {pages} {640} (\bibinfo {year} {2014})}\BibitemShut {NoStop}%
\bibitem [{Note2()}]{Note2}%
	\BibitemOpen
	\bibinfo {note} {The topological defect (monopole) density in spin ice is not in a simple way related to the Kibble-Zurek defect density in the Ising order parameter.  The former naturally relates to the magnetization, with an approximate proportionality close to the critical point, leading to  the same scaling relations}\BibitemShut
	{NoStop}%
\bibitem [{\citenamefont {Tomasello}\ \emph {et~al.}(2015)\citenamefont
		{Tomasello}, \citenamefont {Castelnovo}, \citenamefont {Moessner},\ 
    and\ \citenamefont {Quintanilla}}]{tomasello_single-ion_2015}%
   \BibitemOpen
   \bibfield  {author} {\bibinfo {author} {\bibfnamefont {B.}~\bibnamefont
	  {Tomasello}}, \bibinfo {author} {\bibfnamefont {C.}~\bibnamefont
	  {Castelnovo}}, \bibinfo {author} {\bibfnamefont {R.}~\bibnamefont
	  {Moessner}}, \ and\ \bibinfo {author} {\bibfnamefont {J.}~\bibnamefont
   {Quintanilla}},\ }\href {http://arxiv.org/abs/1506.02672} {\bibfield
	 {journal} {\bibinfo  {journal} {arXiv:1506.02672}}}\BibitemShut {NoStop}%
\bibitem [{\citenamefont {Jaubert}\ and\ \citenamefont
  {Holdsworth}(2011)}]{jaubert_magnetic_2011}%
  \BibitemOpen
  \bibfield  {author} {\bibinfo {author} {\bibfnamefont {L.~D.~C.}\
  \bibnamefont {Jaubert}}\ and\ \bibinfo {author} {\bibfnamefont {P.~C.~W.}\
  \bibnamefont {Holdsworth}},\ }\Doi {10.1088/0953-8984/23/16/164222}
  {\bibfield  {journal} {\bibinfo  {journal} {J.~Phys.: Cond.~Mat.},\ }\textbf
  {\bibinfo {volume} {23}},\ \bibinfo {pages} {164222} (\bibinfo {year}
  {2011})}\BibitemShut {NoStop}%
\bibitem [{\citenamefont {Leeuw}\ \emph {et~al.}(1980)\citenamefont {Leeuw},
  \citenamefont {Perram},\ and\ \citenamefont {Smith}}]{leeuw_simulation_1980}%
  \BibitemOpen
  \bibfield  {author} {\bibinfo {author} {\bibfnamefont {S.~W.~de}\
  \bibnamefont {Leeuw}}, \bibinfo {author} {\bibfnamefont {J.~W.}\ \bibnamefont
  {Perram}}, \ and\ \bibinfo {author} {\bibfnamefont {E.~R.}\ \bibnamefont
  {Smith}},\ }\Doi {10.1098/rspa.1980.0135} {\bibfield  {journal} {\bibinfo
  {journal} {Proc.~R.~Soc.~Lond.~A},\ }\textbf {\bibinfo {volume} {373}},\
  \bibinfo {pages} {27} (\bibinfo {year} {1980})}\BibitemShut {NoStop}%
\bibitem [{\citenamefont {Melko}\ and\ \citenamefont
  {Gingras}(2004)}]{melko_monte_2004}%
  \BibitemOpen
  \bibfield  {author} {\bibinfo {author} {\bibfnamefont {R.~G.}\ \bibnamefont
  {Melko}}\ and\ \bibinfo {author} {\bibfnamefont {M.~J.~P.}\ \bibnamefont
  {Gingras}},\ }\Doi {10.1088/0953-8984/16/43/R02} {\bibfield  {journal}
  {\bibinfo  {journal} {J.~Phys.: Cond.~Mat.},\ }\textbf {\bibinfo {volume}
  {16}},\ \bibinfo {pages} {R1277} (\bibinfo {year} {2004})}\BibitemShut
  {NoStop}%
\bibitem [{\citenamefont {Binder}(1981)}]{binder_finite_1981}%
  \BibitemOpen
  \bibfield  {author} {\bibinfo {author} {\bibfnamefont {K.}~\bibnamefont
  {Binder}},\ }\Doi {10.1007/BF01293604} {\bibfield  {journal} {\bibinfo
  {journal} {Z.~Phys.~B.~Cond.~Mat.},\ }\textbf {\bibinfo {volume} {43}},\
  \bibinfo {pages} {119} (\bibinfo {year} {1981})}\BibitemShut {NoStop}%
\bibitem [{\citenamefont {Schulte}\ and\ \citenamefont
  {Drope}(2005)}]{Schulte2005}%
  \BibitemOpen
  \bibfield  {author} {\bibinfo {author} {\bibfnamefont {M.}~\bibnamefont
  {Schulte}}\ and\ \bibinfo {author} {\bibfnamefont {C.}~\bibnamefont
  {Drope}},\ }\Doi {10.1142/S0129183105007844} {\bibfield  {journal} {\bibinfo
  {journal} {Int.~J.~Mod.~Phys.~C},\ }\textbf {\bibinfo {volume} {16}},\
  \bibinfo {pages} {1217} (\bibinfo {year} {2005})}\BibitemShut {NoStop}%
\bibitem [{\citenamefont {Fenz}\ \emph {et~al.}(2007)\citenamefont {Fenz},
  \citenamefont {Folk}, \citenamefont {Mryglod},\ and\ \citenamefont
  {Omelyan}}]{Fenz2007}%
  \BibitemOpen
  \bibfield  {author} {\bibinfo {author} {\bibfnamefont {W.}~\bibnamefont
  {Fenz}}, \bibinfo {author} {\bibfnamefont {R.}~\bibnamefont {Folk}}, \bibinfo
  {author} {\bibfnamefont {I.~M.}\ \bibnamefont {Mryglod}}, \ and\ \bibinfo
  {author} {\bibfnamefont {I.~P.}\ \bibnamefont {Omelyan}},\ }\Doi
  {10.1103/PhysRevE.75.061504} {\bibfield  {journal} {\bibinfo  {journal}
  {Phys.~Rev.~E},\ }\textbf {\bibinfo {volume} {75}},\ \bibinfo {pages}
  {061504} (\bibinfo {year} {2007})}\BibitemShut {NoStop}%
\bibitem [{\citenamefont {De~Grandi}\ \emph {et~al.}(2010)\citenamefont
  {De~Grandi}, \citenamefont {Gritsev},\ and\ \citenamefont
  {Polkovnikov}}]{de_grandi_quench_2010}%
  \BibitemOpen
  \bibfield  {author} {\bibinfo {author} {\bibfnamefont {C.}~\bibnamefont
  {De~Grandi}}, \bibinfo {author} {\bibfnamefont {V.}~\bibnamefont {Gritsev}},
  \ and\ \bibinfo {author} {\bibfnamefont {A.}~\bibnamefont {Polkovnikov}},\
  }\Doi {10.1103/PhysRevB.81.012303} {\bibfield  {journal} {\bibinfo  {journal}
  {Phys.~Rev.~B},\ }\textbf {\bibinfo {volume} {81}},\ \bibinfo {pages}
  {012303} (\bibinfo {year} {2010})}\BibitemShut {NoStop}%
\bibitem [{\citenamefont {Y~Huang}\ \emph {et~al.}(2014)\citenamefont{Huang}, \citenamefont
  {Yin}, \citenamefont {Feng},\ and\ \citenamefont
  {Zhong}}]{huang_kibble-zurek_2014}%
  \BibitemOpen
  \bibfield  {author} {\bibinfo {author} {\bibfnamefont {Y.}~\bibnamefont
		{Huang}}, \bibinfo {author} {\bibfnamefont {S.}~\bibnamefont {Yin}}, \bibinfo {author} {\bibnamefont {B.}~\bibnamefont{Feng}},\ and\ \bibinfo {author} {\bibfnamefont {F.}~\bibnamefont {Zhong}},\
  }\Doi {10.1103/PhysRevB.90.134108} {\bibfield  {journal} {\bibinfo  {journal}
  {Phys.~Rev.~B},\ }\textbf {\bibinfo {volume} {90}},\ \bibinfo {pages}
  {134108} (\bibinfo {year} {2014})}\BibitemShut {NoStop}%
\bibitem [{\citenamefont {Huang}\ \emph {et~al.}(2015)\citenamefont {Huang},
  \citenamefont {Yin}, \citenamefont {Hu},\ and\ \citenamefont
  {Zhong}}]{huang_scaling_2015}%
  \BibitemOpen
  \bibfield  {author} {\bibinfo {author} {\bibfnamefont {Y.}~\bibnamefont
  {Huang}}, \bibinfo {author} {\bibfnamefont {S.}~\bibnamefont {Yin}}, \bibinfo
  {author} {\bibfnamefont {Q.}~\bibnamefont {Hu}}, \ and\ \bibinfo {author}
  {\bibfnamefont {F.}~\bibnamefont {Zhong}},\ }\href
  {http://arxiv.org/abs/1503.02762} {\bibfield  {journal} {\bibinfo  {journal}
  {arXiv:1503.02762}}}\BibitemShut
  {NoStop}%
\end{thebibliography}
\end{document}